\documentclass[11pt,twocolumn,twoside]{article}
\usepackage{iccr2024}

\usepackage[export]{adjustbox}

\begin{document}

\title{AutoMRISimQA: an automated system for daily quality control of a 3T MRI simulator } 

\author[1,2,3]{Aitang Xing}
\author[1,2,3]{Gary Goozee}
\author[1,2]{Gary Liney}
\author[1,2,3]{Sankar Arumugam}
\author[1,2,3]{Shrikant Deshpande}
\author[1,2,3]{Anthony Espinoza}
\author[1,2,3]{Alison Gray}
\author[1]{Vasilis Kondilis}
\author[1,2]{Doaa Elwadia}
\author[1,2]{Robba Rai }
\author[1,2,3]{Lois Holloway}

\affil[1]{Liverpool \& Macarthur Cancer Therapy Centre, Liverpool, NSW 1871, Australia}

\affil[2]{Ingham Institute for Applied Medical Research, Liverpool, NSW 1871, Australia}

\affil[3]{South West Sydney Clinical Campuses, University of New South Wales, Sydney, Australia}

\maketitle
\thispagestyle{fancy}


\begin{customabstract}
A software system named AutoMRISimQA was developed to monitor the daily performance of a wide-bore 3T scanner (MRI) which was designed and dedicated to radiotherapy simulation. AutoMRISimQA can monitor the performance of the MRI simulator not only by using image quality indices such as signal-to-noise ratio (SNR), uniformity, ghosting, and contrast but also  performing a quick check of geometric accuracy as well as the external lasers quantitatively. It was implemented into the daily clinical workflow in 2013 and has been used for more than 10 years. It was also seamlessly integrated with QATrack+, allowing continuous monitoring of the consistency of the MRI simulator’s performance.
\end{customabstract}


\section{Introduction}
MRI simulator (MRISim) is a magnetic resonant imaging (MRI) scanner designed and dedicated for radiotherapy planning, dose calculation and treatment response \cite{christiansen2016magnetic,king2016functional}. The use of MRI in radiotherapy requires the scanner to be more stable and reproducible than the one used for diagnostic purposes. A recently published TG 284 report recommended a list of tests to be done daily, but how to perform these checks depends on the individual centre \cite{glide2021task}. 

MRISim was commissioned and integrated into the routine workflow in 2013 in our centre \cite{xing2016commissioning,xing2015quality}. A fully automated daily quality control (QC) program was developed and has been running for more than 10 years. The purpose of this paper is to report how the system was designed, implemented and integrated into a radiation oncology workflow.

\section{Materials and Methods}

\subsection{MRI Simulator and phantom}
\begin{figure}
  \centering
  \includegraphics[width=0.45\textwidth,height=0.40\textwidth]{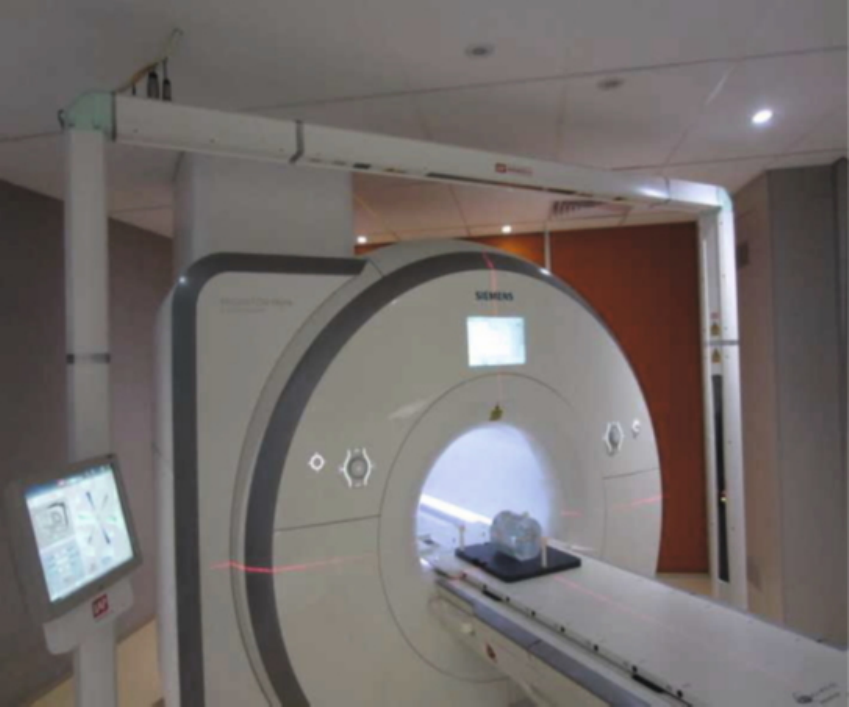}
  \caption{Siemens Skyra 3T MRI Simulator with a flat couch top. The AQUARIUS phantom was set up on the couch and aligned with external lasers. }
  \label{fig:simulator}
\end{figure}

MRI simulator is a wide-bore 3T MRI scanner (Siemens 3T Skyra, Erlangen, Germany) equipped with an external simulation laser and a flat couch \cite{xing2015quality}. The phantom is the AQUARIUS phantom (MRI-0005, S/N 0010260016, LAP of America, Laser application, USA), which was originally designed for checking the alignment of simulation lasers \cite{xing2016commissioning}.  The phantom is cylindrical, and inside the phantom, there are four long tubes filled with CuSO4 solution and the rest is filled with water. As shown in Figure \ref{fig:simulator}, the phantom was set on the flat table aligned with external lasers and then scanned using a T1 weighted spin echo sequence (echo time(TE) =30ms and repetition time(TR)=3410ms) daily. The MRI images were transferred to a shared network location on a server. An in-house system, named AutoMRISimQA was developed to monitor the MRI simulator’s daily performance using phantom images.

\subsection{Design of AutoMRISimQA system }

\begin{figure}
  \flushleft
  \includegraphics[width=0.45\textwidth,height=0.27\textwidth]{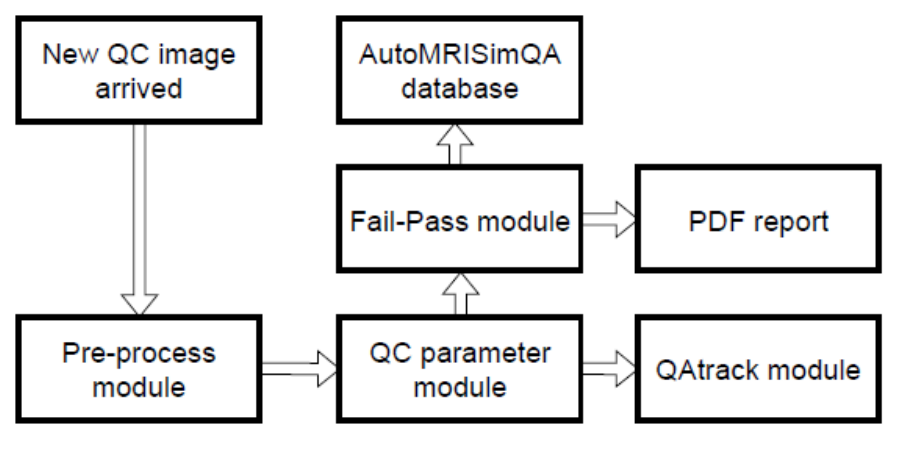}
  \caption{The modular structure of AutoMRISimQA }
  \label{fig:module}
\end{figure}

The system was designed and divided into six modules. Figure \ref{fig:module} showes the overall structure of the system. The first module is used to detect if a new image series arrives. If the new images were found, the second module will be triggered to sort image series, according to the acquisition date and time. The image pixel data were also extracted. The second module uses these data to calculate image quality indices such as signal-to-noise ratio(SNR), which characterize the performance of the MRI scanner. The calculated QC parameters are passed to the third module, whose role is to make a logical judgement. The parameters are compared with their corresponding reference values and tolerance. If any of the QC parameters are out of tolerance, an alert e-mail will be automatically sent to the responsible physicists. A PDF report was also generated at the same time to assist physicists in quickly diagnosing where the problem is.  At our centre, all QC data for radiotherapy equipment were managed using QATrack+ \cite{studinski2014t}, an open-source web-based system designed for radiotherapy quality control. The QATrack module is used to automatically update the database of QATrack+.

\subsection{Implementation of the system }
Matlab, Java and Python were used for system implementation. As QATrack+ was implemented in pure Python, the module for updating the QATrack+ database was written using Python. The access to the database was through the Python representational state transfer(REST)framework \cite{pyrest}. In this way, the risk of accidental damage to the database was minimized. Except for the report module, the rest of the system was implemented using Matlab.  A rich of imaging process functions provided by Matlab’s image processing toolbox is the major consideration of using Matlab as the main programming language. Matlab also can seamlessly integrate Java applications into the Matlab program. The report module was written in Java using the iText package \cite{itext}and integrated into Matlab. 

\begin{figure}
  \flushleft
  \includegraphics[width=0.45\textwidth,height=0.22\textwidth]{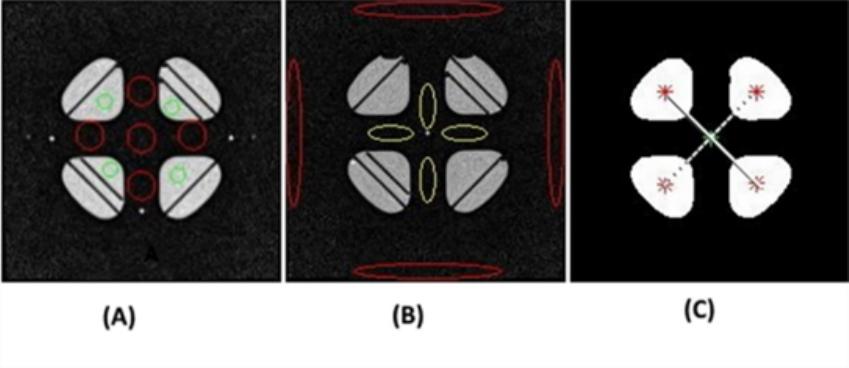}
  \caption{The centre slice of phantom images and ROIs used to calculate QC parameters. }
  \label{fig:phantom}
\end{figure}

The core component of the system is the one used for calculating the QC parameters from the phantom images. The image slice shown in Figure \ref{fig:phantom} (A) was used to calculate the uniformity and SNR using the four \(1cm^{2}\) red circle regions. The percentage intensity uniformity (PIU) is defined as 
\begin{equation}
PIU=100\times (1-\frac{S_{max}-S_{min}}{S_{max}+S_{min}}) 
\label{eq:piu}
\end{equation}

Where \(S_{max}\) and \(S_{min}\) represent the maximum and minimum signal averaged over four red circles.

The SNR is calculated as the ratio of the mean signal averaged over four areas defined by the red circles to the standard deviation over the same ROI. Another parameter, termed output (MRI signal), is defined as the ratio of the standard deviation to the mean of the pixel values over the red circle ROI at the centre of the images. 

The ghosting effect was calculated using the image slice and eight elliptical regions as indicated in Figure \ref{fig:phantom}(B). The ghosting ratio (GR) was calculated as:
\begin{equation}
GR=\frac{\left | \left ( \left ( S_{top}+S_{btm} \right )-\left ( S_{top}+S_{btm} \right ) \right ) \right |}{2S_{mean}}  
\label{eq:gr}
\end{equation}           
Where \(S_{top}\) , \(S_{btm}\)  , \(S_{left} \) and \(S_{right}\) are the signals averaged large eclipses at the top, bottom, right and left of the images, respectively. \(S_{mean}\) is the mean signal averaged over four small ellipsoids at the center of the image.

The contrast was calculated as the ratio of the mean signal value over areas defined by the four green circles in high contrast regions to the mean signal value determined over the four red circle regions. The image was segmented to isolate the four high signal regions and then the centroids of these regions were found. Two diagonal distances between these centers were calculated as shown in Figure \ref{fig:phantom}(C). The forward and backward diagonal distance was recorded as d45 and d135 for reporting purposes.  The laser coordinates Y is the slice position, where laser coordinates X and Z are the center of the central image slice, which were calculated using centroid coordinates of four high contrast regions. 

\begin{figure*}[h!]
  \centering
  \includegraphics[width=0.85\textwidth,height=0.46\textwidth]{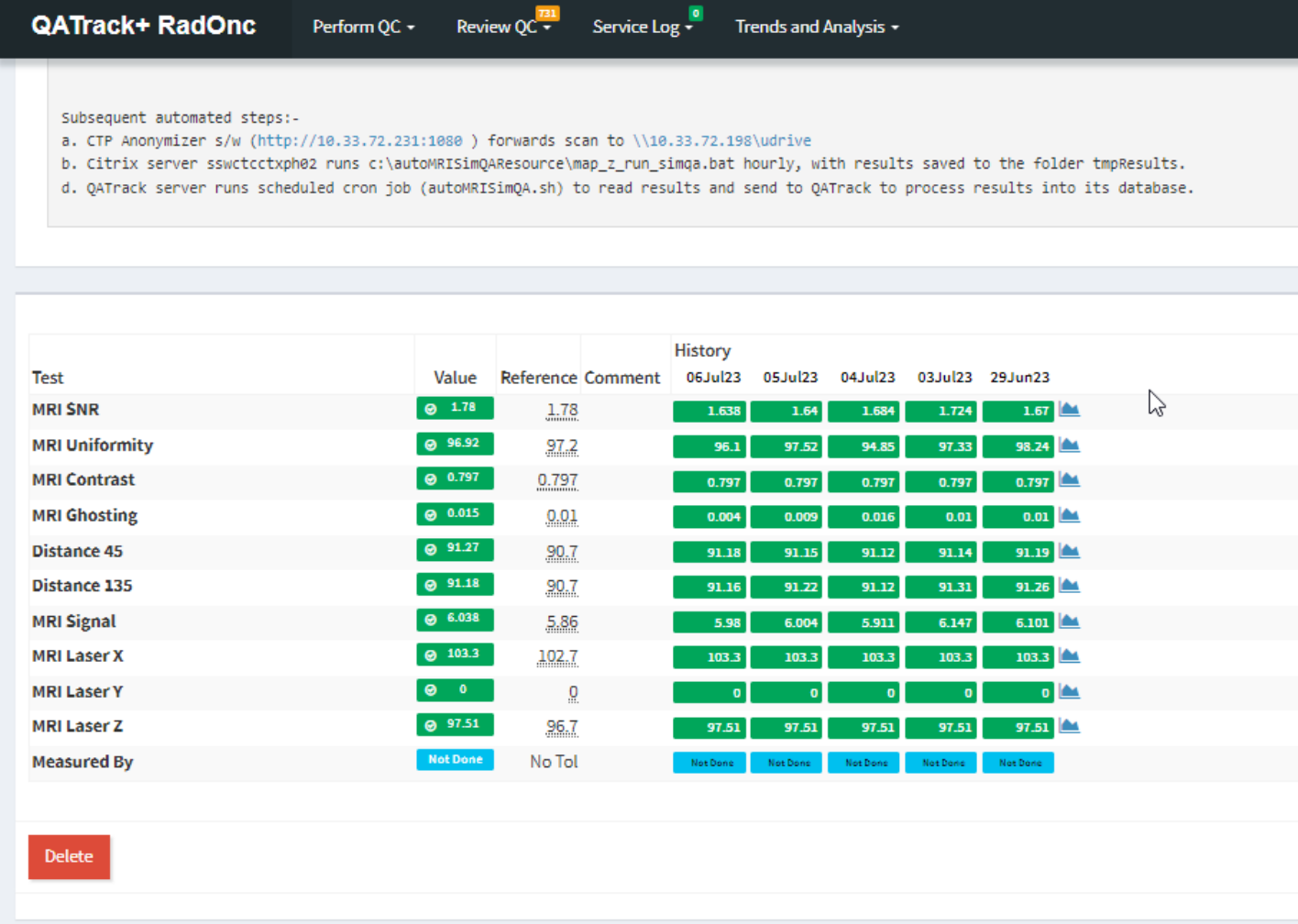}
  \caption{Updated daily QC results displayed through the QATrack+ web interface.}
  \label{fig:qatrack_results}
\end{figure*}

\begin{figure*}[h!]
  \centering
  \includegraphics[width=0.60\textwidth,height=0.35\textwidth]{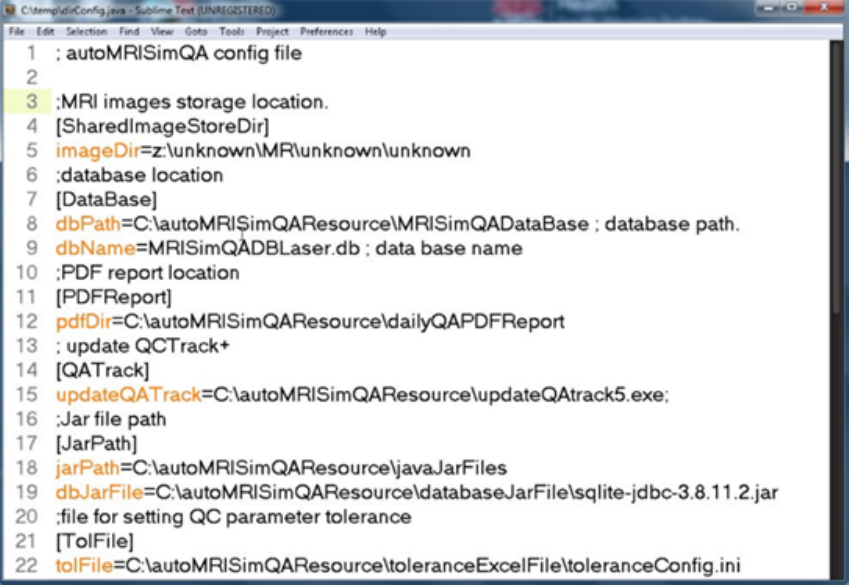}
  \caption{Configurability of the system using a simple text file.}
  \label{fig:settings}
\end{figure*}

\begin{figure*}[h!]
  \centering
  \includegraphics[width=0.85\textwidth,height=0.46\textwidth]{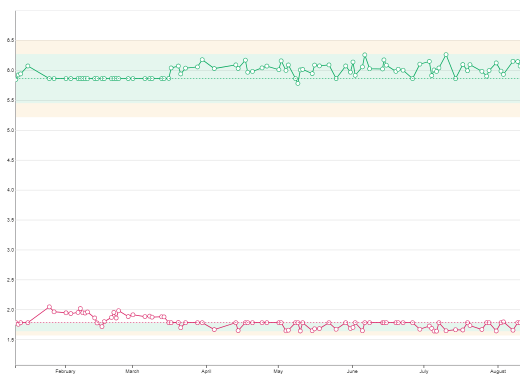}
  \caption{Continuous monitoring of MRI simulator using autoMRISimQA and QATrack+ web interface. SNR (green curve) and MRI signal (red curve) were generated by QATrack+.}
  \label{fig:trend}
\end{figure*}

\section{Results}

The system has been implemented into the MRI-based radiotherapy workflow since 2013. As a start of a daily routine, a radiographer or radiation therapist (RT) scans the laser phantom. Then the images are sent to a location sitting on a shared network drive. The system automatically processes the images and updates the QATrack+ database. The daily QC results can be viewed on any computer connected to the hospital’s private network using the web-based interface of QATrack+ or the PDF report. 

As an example, Figure \ref{fig:qatrack_results} showed one QC result generated by autoMRISimQA results. The traffic lights were used to indicate the acceptability of the results along with the tolerance values. 

The main setting for the system can also be configured via a text file as shown in Figure \ref{fig:settings}. The use of configuration files makes the system more flexible and maintainable.

The statistical analysis based on the data acquired over a certain period provided useful information regarding the MRI’s performance and a tool for continuous monitoring of MRI scanner performance.  As an example, Figure  \ref{fig:trend} shows the trend of SNR and MRI signals over a few months, which were generated by QATrack+.

\section{Discussion}

AutoMRISimQA achieved an automated constancy check of MRISim's performance by monitoring MRI imaging system in one go using multiple image quality indices, which are directly related to the use of MRI in radiotherapy planning and recommended by TG 284 [3], but also doing a quick check of geometric accuracy.  In addition to the visual check of simulation laser alignment, the laser positions were also quantitatively monitored.

The work presented here demonstrates a feasible solution for developing an automated daily QC program for an MRI simulator, which can be used and implemented by other centres.  The source code of the system is available on the request and will be on GitHub soon. Although the system implemented is based on the laser phantom, other commercial phantoms or custom-made phantoms can be used. 

Several authors developed software tools for quality assurance of a diagnostic MRI scanner using ACR MRI phantom \cite{sun2015open,yung2015tu}. An open-source image processing tool, IQWorks, was also used for MRI quality assurance purposes \cite{fazakerley2016application}. In comparison with these tools, autoMRISimQA has several unique features: (1) the system was fully automated once the phantom images were transferred to the network location from the MRI control console; (2) It was seamlessly integrated with an open-source QC data management system. The daily QA results and the trend of the system’s performance can be viewed on any computer connected to the hospital network; (3) Automatic email notification allows physicists monitor the MRISim and take action only when it is required.

The clinical benefit of implementing the AutoMRISimQA system into the clinical workflow is efficiency, especially for a busy clinic with a large number of patients going through the scanner. Before the use of this system, the RTs needed to do two QC tests daily: a customer QC using a uniform phantom and the check of the alignment of simulation lasers using the laser phantom. The average time spent on morning daily QC was half an hour. Now it was reduced to four minutes, the scanning time of laser phantom. Furthermore, the continuous monitoring of MRISim’s performance ensures the short, medium and, especially, long-term follow-up of technical characteristics and their possible drift with time.

\section{Conclusion}

A software system was designed, and implemented for daily quality controls of an MRI simulator. As an open-source tool, the system provides a prototype solution for monitoring the performance of a dedicated MRI scanner used in a radiation oncology department.

\printbibliography

@article{christiansen2016magnetic,
  title={Magnetic Resonance--Only Workflow and Validation of Magnetic Resonance--Based Dose Calculations for Radiation Therapy of Prostate Cancer},
  author={Christiansen, RL and Jensen, HR and Brink, C},
  journal={International Journal of Radiation Oncology, Biology, Physics},
  volume={96},
  number={2},
  pages={E645},
  year={2016},
  publisher={Elsevier}
}

@article{king2016functional,
  title={Functional MRI for the prediction of treatment response in head and neck squamous cell carcinoma: potential and limitations},
  author={King, Ann D and Thoeny, Harriet C},
  journal={Cancer imaging},
  volume={16},
  pages={1--8},
  year={2016},
  publisher={Springer}
}

@article{glide2021task,
  title={Task group 284 report: magnetic resonance imaging simulation in radiotherapy: considerations for clinical implementation, optimization, and quality assurance},
  author={Glide-Hurst, Carri K and Paulson, Eric S and McGee, Kiaran and Tyagi, Neelam and Hu, Yanle and Balter, James and Bayouth, John},
  journal={Medical physics},
  volume={48},
  number={7},
  pages={e636--e670},
  year={2021},
  publisher={Wiley Online Library}
}

@article{xing2016commissioning,
  title={Commissioning and quality control of a dedicated wide bore 3T MRI simulator for radiotherapy planning},
  author={Xing, Aitang and Holloway, Lois and Arumugam, Sankar and Walker, Amy and Rai, Robba and Juresic, Ewa and Cassapi, Lynette and Goozee, Gary and Liney, Gary},
  journal={Int J Cancer Ther Oncol},
  volume={4},
  number={2},
  pages={1--10},
  year={2016}
}

@inproceedings{xing2015quality,
  title={Quality assurance of the radiotherapy workflow integrating a dedicated wide-bore 3T MRI simulator},
  author={Xing, Aitang and Liney, Gary P and Holloway, L and Arumugam, Sankar and Rai, R and Juresic, E and Goozee, Gary},
  booktitle={World Congress on Medical Physics and Biomedical Engineering, June 7-12, 2015, Toronto, Canada},
  pages={539--542},
  year={2015},
  organization={Springer}
}

@article{studinski2014t,
  title={SU-E-T-103: Development and Implementation of Web Based Quality Control Software},
  author={Studinski, R and Taylor, R and Angers, C and La Russa, D and Clark, B},
  journal={Medical Physics},
  volume={41},
  number={6Part12},
  pages={246--246},
  year={2014},
  publisher={Wiley Online Library}
}

@misc{pyrest,
  title = { Python REST API Framework},
  howpublished = {\url{http://python-eve.org/}},
  note = {Accessed: 2024-1-1}
}

@misc{itext,
  title = { Java iText package},
  howpublished = {\url{https://itextpdf.com/}},
  note = {Accessed: 2024-1-1}
}

@article{sun2015open,
  title={An open source automatic quality assurance (OSAQA) tool for the ACR MRI phantom},
  author={Sun, Jidi and Barnes, Michael and Dowling, Jason and Menk, Fred and Stanwell, Peter and Greer, Peter B},
  journal={Australasian physical \& engineering sciences in medicine},
  volume={38},
  pages={39--46},
  year={2015},
  publisher={Springer}
}

@article{yung2015tu,
  title={TU-F-CAMPUS-I-05: Semi-Automated, Open Source MRI Quality Assurance and Quality Control Program for Multi-Unit Institution},
  author={Yung, J and Stefan, W and Reeve, D and Stafford, RJ},
  journal={Medical Physics},
  volume={42},
  number={6Part36},
  pages={3647--3647},
  year={2015},
  publisher={Wiley Online Library}
}

@article{fazakerley2016application,
  title={The application of IQWorks image analysis software to quality assurance in MR imaging},
  author={Fazakerley, Jason and Brunt, JNH and Graham, James and Reilly, Andrew and Moores, BM},
  journal={Biomedical Physics \& Engineering Express},
  volume={2},
  number={2},
  pages={025011},
  year={2016},
  publisher={IOP Publishing}
}

\end{document}